# An option of « UCN pump » for ESS

V.V. Nesvizhevsky

*Institut Max von Laue – Paul Langevin, 71 av. des Martyrs, 38000 Grenoble, France*

**Abstract**

The aim of this short note is to present an option for a source of ultracold neutrons (UCNs), which could profit from the pulse time-structure of the future ESS spallation neutron source in Lund, and thus which could produce a very high UCN density and a rather high UCN flux simultaneously. In order to realize this idea one has to install a relatively thin solid-deuterium UCN source in a close vicinity to the spallation target and to couple it with an extraction UCN guide with an entrance membrane window, which is moving periodically and synchronously with the operation cycle of the spallation source, as explained in the text below. This proposal profits from the fact that all characteristic parameters of the problem, such as the pulse duration of the ESS spallation source, the typical thickness of solid deuterium source that could be easily realized, the typical time of generation of UCNs in solid deuterium, the length and diameter of the extraction neutron guide and the time diagram of the membrane motion that is still realistic, they all fit nicely to optimum desired parameters. The UCN density produced in such a way could approach $10^6$ UCN/cm$^3$.

Ultracold neutrons (UCNs) (1), (2), (3) are an excellent tool in fundamental particle physics (4), (5), (6) and also could potentially be used in the future for typical neutron-scattering applications if much larger UCN densities would be available. Serious efforts to increase UCN densities have been undertaken worldwide basing essentially on two approaches: down-scattering of cold neutrons (mainly with the wavelength of 8.9 Å) in liquid $^4$He and down-scattering of neutrons with broad initial spectrum in solid deuterium (7).

While a characteristic time needed to accumulate a maximum UCN density in sufficiently cold liquid $^4$He is comparable to the neutron lifetime of nearly $10^3$ s and thus only the mean neutron flux is relevant in this case, a characteristic time of production of UCNs in cold solid deuterium (8), (9), (10), (11), (12), (13), (14) is much shorter thus providing us a chance to profit from the pulse structure of neutron flux in a spallation source (15), (16). An additional evident advantage of using spallation sources over nuclear reactors for UCN production in solid deuterium consists of significantly lower γ-heat load per equivalent initial neutron flux. Therefore we leave aside here the options for liquid-helium UCN sources at ESS, which could be implemented differently and are (will be) considered in other publications, and focus only on solid deuterium.

In order to produce a maximum UCN density in solid deuterium, a UCN converter should be placed in the vicinity of the spallation target, for example in the through-going tube which is currently being discussed at the ESS (17). As the optimum temperature of incident neutrons is around 40K (12), a liquid-hydrogen or liquid-deuterium (pre)moderator would be useful to cool neutrons produced in the spallation target, in order to increase the UCN output. Without going into details of relevant estimations, we could refer to the available numerous estimations and measurements of UCN production in analogous conditions as mentioned in the above references (corrected for the value of the pulse neutron flux at the ESS): the UCN density could reach over $10^6$ UCN/cm$^3$ at optimum conditions.

For simplicity, imagine a solid-deuterium disk with the diameter equal to that of the liquid-hydrogen or –deuterium (pre)moderator, and the thickness equal to, say, half a centimeter (0.5 cm). During a pulse of the spallation source equal to 2.5 ms, UCNs would be generated in the whole body of solid deuterium up to their saturation density. With the typical total velocity of 4-5 m/s and isotropic angular distribution, UCNs would also fill the layer with the thickness of about 0.5 cm outside the source bulk (let's call it a "halo" around the source) with the density equal about a half of the maximum saturation density (no significant losses in solid deuterium at such a small deuterium thicknesses; a mirror for UCNs on the back side of the deuterium disk would be useful in order to provide effectively the double thickness of solid deuterium). A design of such a thin solid-deuterium source, which is well suited for our application, is considered in (11).

Is it possible to extract such a high "peak" UCN density from the halo to an experimental setup outside the biological shielding with limited losses in UCN phase-space density?

Yes, this could be done using the idea of "UCN pump" (18) discussed here. It is based on the fact that UCNs are reflected from a membrane at rest (or if its velocity is significantly lower than its critical velocity); however UCNs penetrate through the same membrane with a high efficiency if it is moving towards UCNs with a velocity significantly larger than its critical velocity.

Imagine a thin membrane, which could move periodically through the halo in such a manner that it moves slowly away from the deuterium disk when the spallation source is "off" (~60 ms, 0.5 cm, ~0.1 m/s), and rapidly approaches to it when the halo zone is filled in with UCNs (0.5 ms, 0.5 cm, ~10 m/s). During the motion away from the deuterium disk, the velocity of membrane is small enough, so nearly all UCNs (trapped in the extraction UCN guide) are reflected from it. During the motion towards the deuterium disk, the velocity of membrane is large enough to make it nearly transparent for UCNs (being in the "halo"); note that UCNs in the extraction guide move slower than the membrane, and thus they even do not touch the membrane at this phase of the membrane motion. Thus the membrane acts as a good "pump". The density in a sufficiently small extraction neutron guide would continue increasing until it approaches the "peak" UCN density in the halo.

There are at least a few practical means for realizing such a membrane. In any case it should be a thin foil (a small mass will facilitate fast oscillating motions of the foil) with high critical energy (to increase the maximum energy of UCNs that could be trapped in the UCN extraction guide and used in experiments outside the spallation source). The key parameters, which define the choice of material for the foil, are mechanical properties providing long-term mechanical resistance and also not too high absorption and inelastic scattering of UCNs during their penetration through the moving foil. The motion could be realized using He-based pneumatic system, or elastic springs based on eccentric rotations, or pièzo-elements.

What is the volume of the neutron guide that could be filled in with such a peak density of UCNs?

For a typical storage time of UCNs in the neutron guide of, say, 60 s, ~$10^3$ pulses of the spallation source would contribute to building up the UCN density, thus giving us an estimation of the guide length of 5 m, filled in with the estimated high UCN density (in case if the guide cross-section is equal to the cross-section of the solid deuterium disk). This length is comparable to the thickness of the biological shielding of the ESS spallation source; if the extraction length has to be longer, a "cheap" method is to decrease accordingly the guide diameter. In order to optimize the overall performance of the UCN source, the diameter and thickness of the deuterium disk, the diameter of the UCN guide and the storage time, the parameters of motion of the membrane and so on should be adjusted accordingly. This optimization is out of the scope of this present short note, but could and should be done.

To summarize: the goal of this note was to propose a new option for a UCN source at ESS and to provide rough estimations, needed for judging preliminary on the feasibility of its realization. On one hand this idea allows you to profit from all principle advantages of the ESS neutron source (long pulse, record neutron peak flux, modest γ-load) (19) for production of UCNs. On another hand its realization is based on known technologies and looks promising although a detailed further analysis, optimization and feasibility tests are needed.

In order to allow for an implementation of such a source, a going-through tube in a close vicinity to the spallation target and (preferably) close to a liquid hydrogen/deuterium source should be previewed in the initial design. One of possible designs of implementation of the "UCN pump" is shown in Fig. 1.

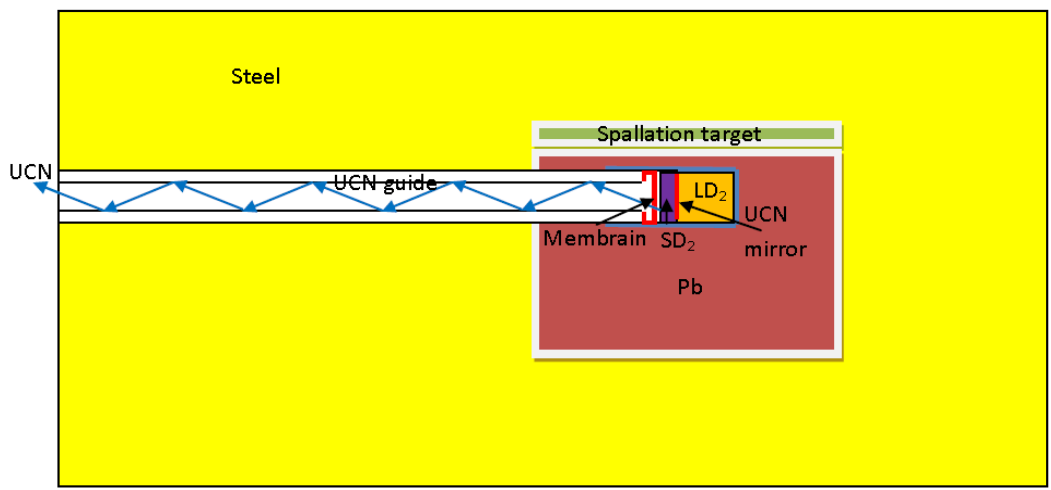

Fig. 1

*A principle scheme of "UCN pump" for the ESS spallation neutron source is illustrated in this figure. Fast neutrons are produces in the spallation target; they are thermalized, for instance, in a liquid deuterium source of cold neutrons to neutron energies close to the optimum for UCN production in a thin solid deuterium UCN source; UCNs are transported through a specular UCN guide to an experimental installation placed outside the biological shielding of the ESS; UCNs could penetrate through a membrane moving in a particular manner: the membrane moves rapidly towards $SD_2$ when the peak UCN density is produced in the halo around $SD_2$ and it moves slowly to the opposite direction at any other times.*

## Travaux cites